\documentclass[fleqn,usenatbib]{mnras}

\usepackage{newtxtext,newtxmath}

\usepackage[T1]{fontenc}

\DeclareRobustCommand{\VAN}[3]{#2}
\let\VANthebibliography\thebibliography
\def\thebibliography{\DeclareRobustCommand{\VAN}[3]{##3}\VANthebibliography}

\usepackage{graphicx}	
\usepackage{amsmath}	

\title[Using MOPED to identify most constraining ice observations]{Identifying the most constraining ice observations to infer molecular binding energies}

\author[J. Heyl et al.]{
Johannes Heyl,$^{1}$\thanks{E-mail: johannes.heyl.19@ucl.ac.uk}
Elena Sellentin,$^{2,3}$
Jonathan Holdship$^{2,1}$
and Serena Viti$^{2,1}$
\\
$^{1}$Department of Physics and Astronomy, University College London, Gower Street, WC1E 6BT, London, UK\\
$^{2}$Leiden Observatory, Leiden University, Huygens Laboratory, Niels Bohrweg 2, NL-2333 CA Leiden, The Netherlands\\
$^{3}$Mathematical Institute, Leiden University, Snellius Building, Niels Bohrweg 1, NL-2333 CA Leiden, The Netherlands
}

\date{Accepted XXX. Received YYY; in original form ZZZ}

\pubyear{2015}

\begin{document}
\label{firstpage}
\pagerange{\pageref{firstpage}--\pageref{lastpage}}
\maketitle

\begin{abstract}
In order to understand grain-surface chemistry, one must have a good understanding of the reaction rate parameters. For diffusion-based reactions, these parameters are binding energies of the reacting species. However, attempts to estimate these values from grain-surface abundances using Bayesian inference are inhibited by a lack of enough sufficiently constraining data. In this work, we use the Massive Optimised Parameter Estimation and Data (MOPED) compression algorithm to determine which species should be prioritised for future ice observations to better constrain molecular binding energies. Using the results from this algorithm, we make recommendations for which species future observations should focus on. 
\end{abstract}



\section{Introduction} \label{sec:intro}
Interstellar dust grains are a crucial component of interstellar chemistry. Many gas-phase complex organic molecules (COMs) have been detected in our galaxy in cold and hot cores \citep{Boogert}. There is evidence to suggest that much of the observed chemistry takes place on the grain surfaces as opposed to the gas-phase and that these observed gas-phase molecules simply evaporate from the grains some time after formation.   As such, if one wishes to understand how such complex organic molecules are formed, one must have a thorough understanding of grain-surface chemistry \citep{Herbst, Caselli}. 

In order to better understand how grain surface chemistry proceeds, it is important to know the reaction rate parameters. For grain-surface reactions, these parameters may not necessarily be the rates themselves, but rather parameters that are more specific to the reaction rate mechanism. For diffusion-based reactions, which are typically taken to be the dominant grain-surface reaction mechanism, the reaction rate parameters of relevance are the binding energies of the reacting species and reaction activation energy barriers \citep{Hasegawa}. Much experimental work has been done to determine these, but there are often significant disagreements, due to differing laboratory conditions (see \cite{Penteado} for a survey of binding energy values). 

There exist a variety of methods to estimate the binding energies, ranging from experimental approaches \citep{experimental_approach} to density functional theory \citep{Ferrero} to machine learning approaches \citep{Villadsen}. However, in our work to estimate these reaction rate parameters given observed abundances, Bayesian inference is typically employed. Bayesian inference has become a ubiquitous tool in astrophysics and has recently found more use within the field of astrochemistry. Previous work has considered the rate-parameter estimation problem \citep{holdship, Heyl1} and has shown that the paucity of available grain-surface species abundances inhibits precise estimates of these rate parameters. The problem due to the lack of sufficiently constraining data has been somewhat ameliorated by considering the network structure \citep{Heyl1} or the underlying chemical mechanisms to reduce the dimensionality of the problem \citep{Heyl2}. However, it remains the case that many binding energies cannot be constrained to the point that they would be useful in chemical codes. This is clear from a survey of the literature which shows quite significant disagreements for some binding energy values \citep{UMIST, Wakelam, Quenard}.

Observations of the ices have typically considered the molecular vibration transitions in the infrared region \citep{Boogert}. A number of space telescopes such as the Infrared Space Observatory (ISO) and Spitzer have provided observations of ice band profiles that have been used to determine molecular abundances. However, until now there has been insufficient resolution of the absorption band profiles. The James Webb Space Telescope (JWST) observes in the infrared wavelength range of 0.6 - 28 $\mu$m. It provides higher spectral resolution observations of up two magnitudes, especially in the 5-8 $\mu$m range which potentially contains the vibrational modes of several molecules of interest \citep{Boogert, Boogert_conference}. This is particularly important as infrared spectroscopy reveals the features of various functional groups which differ by species but can have similar values \citep{Boogert_conference}. As such, having greater resolution will ensure that the various absorption band profiles can be disentangled. 

In this work, we wish to provide recommendations of which species should be prioritised for future ice observations in order to reduce the uncertainties on the binding energy values. To achieve this, we make use of the "Massive Optimised Parameter Estimation and Data compression" (MOPED) algorithm \citep{Heavens, Heavens2017, Heavens2020ExtremePhysics}. A key output of the MOPED algorithm is a measure of how strongly knowledge of a species ice phase abundance would constrain the binding energies.

We start by explaining the chemical code and network we will use throughout this work in Section 2. Section 3 will be dedicated to explaining the approach we take in this work, specifically our use of Bayesian inference and the MOPED algorithm. We follow this up in Section 4 by showing the results of the Bayesian inference and the MOPED algorithm as well as by discussing  the observational implications of our findings. We briefly conclude in Section 5.

\section{The Chemical Code and Network} \label{sec:style}
\subsection{The Chemical Code}

In this work, the gas-grain astrochemical code UCLCHEM \citep{UCLCHEM} was used to model the chemistry of a collapsing dark cloud. The cloud was taken to collapse isothermally at 10K from $10^{2}$ cm$^{-3}$ to $10^{6}$ cm$^{-3}$ over a period of 5 million years. By the end of this collapse, we expect the ice phase abundances to be representative of a dark cloud.  

\subsection{Grain Surface Chemistry}
\subsubsection{Grain Surface Diffusion} It is important to understand the grain surface mechanisms, as this is needed to show why this work considers binding energies as the key parameters that govern the reaction rates. 

We assume all grain surface reactions take place via the Langmuir–Hinshelwood mechanism and use the formalism described in \cite{Hasegawa}  which was implemented in UCLCHEM in \cite{Quenard}. We believe this is a reasonable assumption as previous work has shown that including Eley-Rideal reactions does not strongly affect surface abundances \citep{Ruaud}. According to the formalism, the rate at which two species A and B react via diffusion is given by: 

\begin{equation}\label{reaction_rate}
k_{AB} = \kappa_{AB}\frac{(k^{A}_{hop}+k^{B}_{hop})}{N_{site}n_{dust}},
\end{equation}

where $N_{site}$ is the number of sites on the grain surface and $n_{dust}$ is the dust grain number density.




In equation \ref{reaction_rate}, $k^{X}_{hop}$ is the thermal hopping rate of species $X$ on the grain surface which is defined as: 

\begin{equation}\label{hopping_rate_equation}
k^{X}_{hop} = \nu_{0}\exp\left(-\frac{E_{D}}{T_{gr}}\right),
\end{equation}

where $E_{D}$ is the diffusion energy of the species, $T_{gr}$ is the grain temperature and $\nu_{0}$ is the characteristic vibration frequency of species $X$. The diffusion energy is a fraction of the binding energy of the species, $E_{b}$. In this work, this fraction is taken to be 0.5, in line with \cite{Quenard}. While it is known that this value can vary between 0.3 and 0.8, there is significant uncertainty within that range \citep{GarrodandPauly}. Furthermore, the value is not expected to play a significant role at 10 K \citep{Vasyunin}.

The characteristic vibration frequency, $\nu_{0}$, is defined as: 

\begin{equation}
\nu_{0} = \sqrt{\frac{2k_{b}n_{s}E_{b}}{\pi^{2}m}},
\end{equation}

where $k_{b}$ is the Boltzmann constant, $n_{s}$ is the grain site density and $m$ is the mass of species.While there exists some debate regarding the validity of this expression (see \cite{characteristic_vibration_frequency} for a more detailed discussion), this equation for the characteristic vibration frequency is what is used in UCLCHEM. While a more accurate equation that takes into account the rotation partition function of the desorbing molecules should be used, this will not affect the ability of Bayesian inference to constrain the binding energies of species of interest, which is the aim of this paper.




The final term, $\kappa_{AB}$, which gives the reaction probability is:

\begin{equation}\label{kappa}
\kappa_{AB} = \max\left(\exp{\left(-\frac{2a}{\hbar}\sqrt{2\mu k_{b}E_{A}}\right)}, \exp{\left(-\frac{E_{A}}{T_{gr}}\right)}\right),
\end{equation}

where $\hbar$ is the reduced Planck constant, $\mu$ is the reduced mass, $E_{A}$ is the reaction activation energy, $k_{b}$ is Boltzmann's constant and $a = 1.4$ Angstrom is the thickness of a quantum mechanical barrier. While values between 1 and 2 Angstrom have been used \citep{Hasegawa, GarrodandPauly,Vasyunin}, \cite{Quenard} found that a value of 1.4 Angstrom matched the ice composition best. The reaction probability represents the competition between the quantum mechanical probability of a tunnelling through a rectangular barrier of thickness $a$, which is the first term, and the thermal reaction probability, which is the second term.

\subsubsection{Reaction-diffusion competition}
A modification needs to be made to the  $\kappa_{AB}$ term to take into account the possibility that species might diffuse or evaporate before they can react with each other. This is the reaction-diffusion competition \citep{Chang, GarrodandPauly}. The reaction probability is now defined as: 

\begin{equation}
\kappa_{AB}^{final} = \frac{p_{reac}}{p_{reac} + p_{diff} + p_{evap}},
\end{equation}

where $p_{reac}$, $p_{diff}$ and $p_{evap}$ represent the probabilities of species A and B reacting, diffusing and evaporating per unit time, respectively. These quantities are defined as: 

\begin{equation}
p_{reac} = \max(\nu_{0}^{A}, \nu_{0}^{B})\kappa_{AB}
\end{equation}, 

\begin{equation}
p_{diff} = k_{hop}^{A} + k_{hop}^{B}
\end{equation}

and

\begin{equation}
p_{evap} = \nu_{0}^{A}\exp\left(-\frac{E_{b}^{A}}{T_{gr}}\right) + \nu_{0}^{B}\exp\left(-\frac{E_{b}^{B}}{T_{gr}}\right).
\end{equation}

We replace $\kappa_{AB}$ with $\kappa_{AB}^{final}$ in Equation \ref{reaction_rate},.

Overall, we find that Equations 1-8 show that the key quantities are $\nu_{0}$, $k_{hop}^{X}$, $E_{b}$ and $E_{A}$. The first three are all functions of the binding energies of the reacting species, indicating the binding energies are the crucial parameters. We assume that the activation energies in Equation \ref{kappa} are well-known. This is reasonable, as these should be independent of the ice composition (unlike the binding energies) and can be determined theoretically or experimentally. Many of the reactions would also be expected to have zero activation energy as they are radical-radical reactions \citep{Quenard}. 

\subsection{The Chemical Network}
\label{sec:network}
The chemical network consists of a gas-phase network taken from UMIST12 \citep{UMIST} and a grain-surface network based on \citet{Quenard} and expanded to include the reactions from \citet{Garrod_2008, Minissale_2016, Quan_2010, Fedoseev_2016, Belloche_2017, Song2016,Garrod2006}.

We believe the gas phase network is comprehensive and sufficiently accurate that any deficiencies in the network will not have a great effect on our results. The gas-phase network was benchmarked against observations in \cite{UMIST}. The abundances of species freezing out from the gas phase are likely to be approximately correct and we therefore only need to be concerned by the accuracy and completeness of the grain surface network. We operate under the assumption that the gas-phase network is complete.

Our grain surface network is less comprehensive but we argue it is sufficient to reproduce the abundance of major species, given the results of \cite{Antonios, holdship, Heyl1, Heyl2} which used smaller networks. The network includes the freeze out of all species, hydrogenation reactions of all species up to their saturated forms, and radical-radical reactions that have been shown to be efficient in laboratory experiments, as well as other diffusion reactions from the literature (see above). By including all reactions known to be the main routes through which species like H$_2$O and CH$_3$OH are formed on the grain surfaces, our network is  sufficient to produce accurate ice phase abundances of these species. Therefore, we can properly predict how important the binding energies of those species are to the surface chemistry.

\section{Analytical Approach} \label{sec:displaymath}
\subsection{Parameters}
The aim of this work is to determine the binding energies of the chemically reactive species. While it would be ideal to determine the binding energies of all species in the network, the reality of the situation is that this is not strictly necessary. In \cite{Heyl2}, it was demonstrated that at 10K, a moderate difference in binding energies between two species results in a significant difference in reaction rates. As such, one can significantly reduce the dimensionality of the problem one is trying to solve by only considering the most diffusive species. These are those species that will be the more reactive species with the greater hopping frequency for at least one reaction in the network. The more reactive species were determined by considering the literature. Even though there is widespread disagreement about the values of the binding energies, there is less disagreement about the hierarchy of binding energy values. This can be seen by considering the values given in \cite{Wakelam}, \cite{UMIST} and \cite{Penteado}. For reactions where the literature was not definitive in specifying which species had the lower binding energy, both species' binding energies were included as parameters. The binding energies we considered as parameters were the binding energies of H, H$_{2}$, C, CH, N, CH$_{3}$, NH, CH$_{4}$ and O.

\begin{table}
\setlength\tabcolsep{1.5pt}
\hspace*{1.5cm}
 \begin{tabular}{||c c c||} 
 \hline
 Species & Abundances relative to H  & Source\\ [1ex] 
 \hline\hline
 H$_{2}$O & $(4.0 \pm 1.3) \times 10^{-5}$ &  Cloud \\ 
 \hline
 CO & $(1.2 \pm 0.8) \times 10^{-5}$ &  Cloud\\
 \hline
 CO$_{2}$ & $(1.3 \pm 0.7) \times 10^{-5}$ &  Cloud\\
 \hline
 CH$_{3}$OH & $(5.2 \pm 2.4) \times 10^{-6}$ &  Cloud\\
 \hline
  NH$_{3}$ & $(3.6 \pm 2.6) \times 10^{-6}$ & LYSOs\\
 \hline
 CH$_{4}$ & $(2.3 \pm 2.1) \times 10^{-6}$ & LYSOs\\
 \hline
  HCOOH & $(2.4 \pm 1.3) \times 10^{-6}$ & LYSOs \\
 \hline
  NH$_{4}^{+}$ & $(3.8 \pm 1.5) \times 10^{-6}$ &  Cloud\\
  \hline
\end{tabular}
\caption{The abundances and uncertainties taken for the network adapted from \protect\cite{Boogert}.}
\label{abundance_table}
\end{table}

\subsection{Bayesian Inference}\label{bayesian_inference}

\subsubsection{Introduction to Bayesian Inference}
The goal is to estimate the binding energies of the most diffusive species in this network. We represent these parameters of interest as a vector, \textbf{E} = (E$_{b, H}$, E$_{b,H_{2}}$, E$_{b,C}$, E$_{b,CH}$, E$_{b,N}$, E$_{b,CH_{3}}$, E$_{b, NH}$, E$_{b,CH_{4}}$, E$_{b,O}$). UCLCHEM was modified so that it took these values as an input and output all the final abundances of grain-surface abundances. We represent the 72 grain-surface abundances as a vector $\textbf{Y} = (Y_{1}, Y_{2} ... Y_{72})$. The mapping between $\textbf{E}$ and $\textbf{Y}$ is simply UCLCHEM and we can write this as $\textbf{Y} = f(\textbf{E})$. 

In order to solve the inverse problem, we require abundance measurements of grain-surface species, $\textbf{d}$. These are listed in Table \ref{abundance_table}. These are taken from \cite{Boogert}. 

Bayes' Law can be used to determine the posterior distribution of the binding energies given the data: 

\begin{equation}
P(\textbf{E} \vert \textbf{d}) = \frac{P(\textbf{d} \vert \textbf{E})P(\textbf{E})}{P(\textbf{d})},
\end{equation}
where $P(\textbf{E} \vert \textbf{d})$ is the posterior probability distribution, $P(\textbf{E})$ is the prior, $P(\textbf{d} \vert \textbf{E})$ is the likelihood and $P(\textbf{d})$ is referred to as the evidence. The prior distribution encodes the initial understanding of the binding energy distribution. The likelihood gives the data's likelihood as a function of the binding energies. Within the likelihood function, the physical model is encoded. The evidence serves as a normalising factor and represents the marginalised likelihood. The posterior distribution represents the updated probability distribution of reaction rates based on the data, the prior distribution, and the physical model.

\subsubsection{Implementation}
The prior for all binding energies was specified as uniform distribution between 400 K and 2000 K. The abundance measurements in Table \ref{abundance_table} were assumed to be Gaussian which allowed for the specification of a Gaussian likelihood function: 

\begin{equation}\label{likelihood}
\centering
P(\textbf{d} \vert \textbf{E}) =  \prod_{i=1}^{n_{d}} \frac{1}{\sqrt{2\pi}\sigma_{i}} \exp\left({-\frac{(d_{i}-Y_{i})^{2}}{2\sigma_{i}^{2}}}\right),
\end{equation}

where $n_{d}$ is the number of observations and $\sigma_{i}$ is the uncertainty of the $i$th observation. Only the species for which there are abundances are indexed over. 

The UltraNest Python package \citep{UltraNest} was used for the Bayesian inference, which is based on the MLFriends algorithm \citep{Buchner1, Buchner2}. The package also outputs the maximum likelihood-estimator, $\boldsymbol{E_{ML}}$. We will use this later for the MOPED algorithm.

\subsection{The MOPED Algorithm}
The aim of the MOPED algorithm is to determine which of the $M$ species in our chemical network need to be prioritised for future ice observations in order to best constrain the posteriors for our $p$ parameters. In our situation, $p = 9$ and $M = 72$. In other words, we wish to determine which species will provide us with the most information upon its detection. 

Recall that we wish to determine a set of parameters $\boldsymbol{E}$.  The species that are found to be important may include the species already listed in Table \ref{abundance_table}, in which case we would aim to improve the uncertainties surrounding their values. However, it is also possible that we would need to detect species that have not been detected yet. 

All of our future measurements will be have some instrumental uncertainty. For our purposes, we assume the uncertainty on each measurement will be the same. We define a covariance matrix to summarise this: $\textbf{C} = diag(\sigma_{1}^{2}, \sigma_{2}^{2}, ... \sigma_{M}^{2})$. By operating under this assumption that we can measure any species to the same level of abundance uncertainty, we are aiming to determine which species would be the most useful to detect. In general, it might be the case that different species have different levels of uncertainty.

It is likely that some species will be significantly more impactful in providing information about the parameters of interest. As such, we need to identify the species in question. To this end, we will use a filtering technique developed by \citet{Heavens, Heavens2017, Heavens2020ExtremePhysics} who propose using a linear combination of the final abundances of network, $\boldsymbol{Y}$, to compress data points. Such a compression would be of the form: 

\begin{equation}
    c_{\alpha} = \boldsymbol{b_{\alpha}^{T}}\boldsymbol{Y},
    \label{compressed_data_point}
\end{equation}

where $\alpha$ ranges from 1 to $p$ and ${\boldsymbol{b_{\alpha}}}$ is a set of orthonormal linear filters, such that each one contains as much information about that parameter that is not contained in any other ${\boldsymbol{b_{\alpha}}}$. $\boldsymbol{Y}$ represents a vector containing the final abundances for some arbitrary value of $\textbf{E}$. As a fiducial model, we typically take $\boldsymbol{E} = \boldsymbol{E_{ML}}$, which we can determine using the Bayesian inference discussed in Section \ref{bayesian_inference}. Using the maximum-likelihood parameters as a fiducial model has been found to be sufficient \citep{Heavens,Heavens2017}. The value of each $c_{\alpha}$ will ultimately be more strongly influenced by the components ${\boldsymbol{b_{\alpha}}}$ that are larger in magnitude. As there is one species for each component, this means that if a component has a greater magnitude then it contains more information about that parameter.

The vectors ${\boldsymbol{b_{\alpha}}}$ are given by

\begin{equation}
    \textbf{b}_1 = \frac{\textbf{C}^{-1} \textbf{Y},_1}{\sqrt{\textbf{Y},_1^T \textbf{C}^{-1} \textbf{Y},_1 }}
    \label{filter_1}
\end{equation}

and 

\begin{equation}
    \textbf{b}_\alpha = \frac{\textbf{C}^{-1} \textbf{Y},_\alpha - \sum_{\beta = 1 }^{\alpha -1}{(\textbf{Y,}_\alpha^T \textbf{b,}_\beta)\textbf{b,}_\beta }}{\sqrt{ \textbf{Y,}_\alpha^T \textbf{C}^{-1} \textbf{Y},_\alpha - \sum _{\beta = 1}^{\alpha -1 }(\textbf{Y},_\alpha^T \textbf{b},_\beta)^2}}, 
    \label{filter_2}
\end{equation}

where $\textbf{Y},_{\alpha}$ is the partial derivative of $\textbf{Y}$ with respect to the parameter $\alpha$. The equations for $b_{\alpha}$ were derived in \cite{Heavens} through a Lagrange multiplier procedure. The iterative process of determining each linear filter $\boldsymbol{b_{\alpha}}$ from previous ones is akin to the Gram-Schmidt orthogonalisation. This ensures that all the filters are orthonormal, that is 

\begin{equation}
    \boldsymbol{b_{\alpha}^{T}}\boldsymbol{C}\boldsymbol{b_{\beta}} = \delta_{\alpha\beta},
    \label{orthonormality}
\end{equation}

which is important because it means that all the filter vectors are uncorrelated. Note also that each component of $b_{\alpha}$ is weighted towards species which are low in noise, as measured by the inverse covariance matrix, as well as species with a greater impact on the parameter, as determined by the values in $\textbf{Y},_\alpha$.

Ultimately, we find that vector of abundances of all species $x$ which has dimensionality M has been reduced to $p$ numbers, where $p < M$. This data compression is lossless, which means the same information is included in the $p$ values of $c_{\alpha}$. This was originally stated in \cite{TTH} and proven in \cite{Heavens}. 

Recall that the magnitude of each component of $\boldsymbol{b_{\alpha}}$ gives a weighting for that species' influence on the parameter $\alpha$. To determine the best species to prioritise detection for, we simply add the absolute values of the components of $\boldsymbol{b_{\alpha}}$ for species across all $\alpha$. That is, we perform the sum over our linear filters

\begin{equation}
\sum_{\alpha=1}^{p} [|b_{\alpha}^{1}|, |b_{\alpha}^{2}|..., |b_{\alpha}^{M}|].
\end{equation}

We now have a ``filter sum" for each of the $M$ species in our network. We can rank the species by their filter sum in order to determine which ones have the greatest impact on our parameters.

\section{Results}
\subsection{Results of the Bayesian Inference}

\begin{figure*}
    \includegraphics[width=2.3\columnwidth]{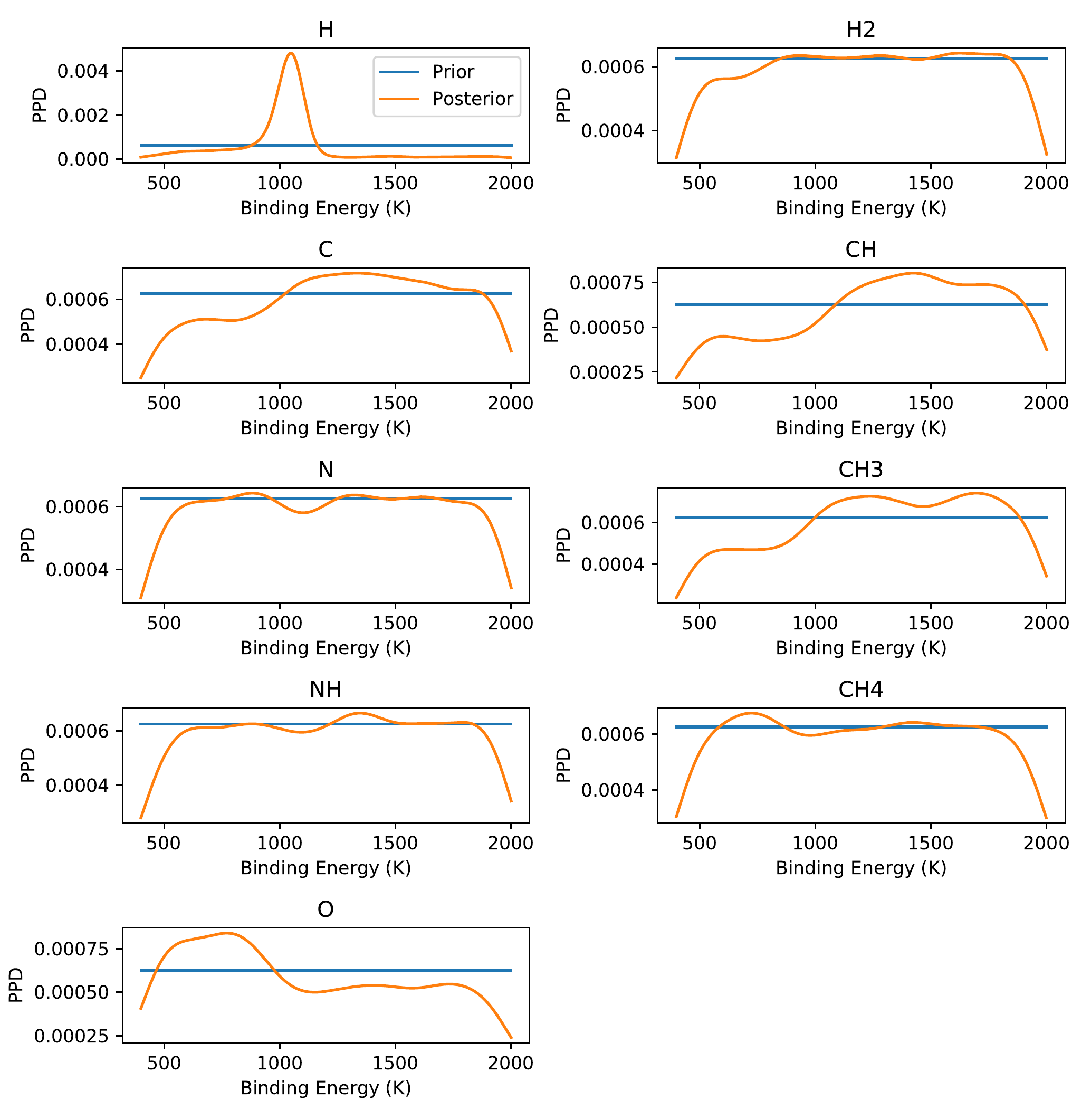}
    \caption{Marginalised posterior distributions of the binding energies of the diffusive species of interest. Also plotted is the prior distribution on the binding energies. With the exception of H, most binding energy distributions differ very little from the prior distribution. This is due to the lack of enough sufficiently constraining data. This motivates the need for further ice observations to reduce the variance of the distributions. }
    \label{binding_energy_plot}
\end{figure*}

Figure \ref{binding_energy_plot} shows the marginalised posterior distributions for the binding energies of interest. The marginalised prior distribution is also plotted for comparison. It is clear that, with the exception of atomic hydrogen's binding energy, the marginalised posterior distributions differ very little from the prior suggesting a lack of sufficiently constraining data. It is for this reason that we now use the MOPED algorithm to identify species we need to detect to better constrain our posterior distributions. 

\begin{figure*}
    \includegraphics[width=2.3\columnwidth]{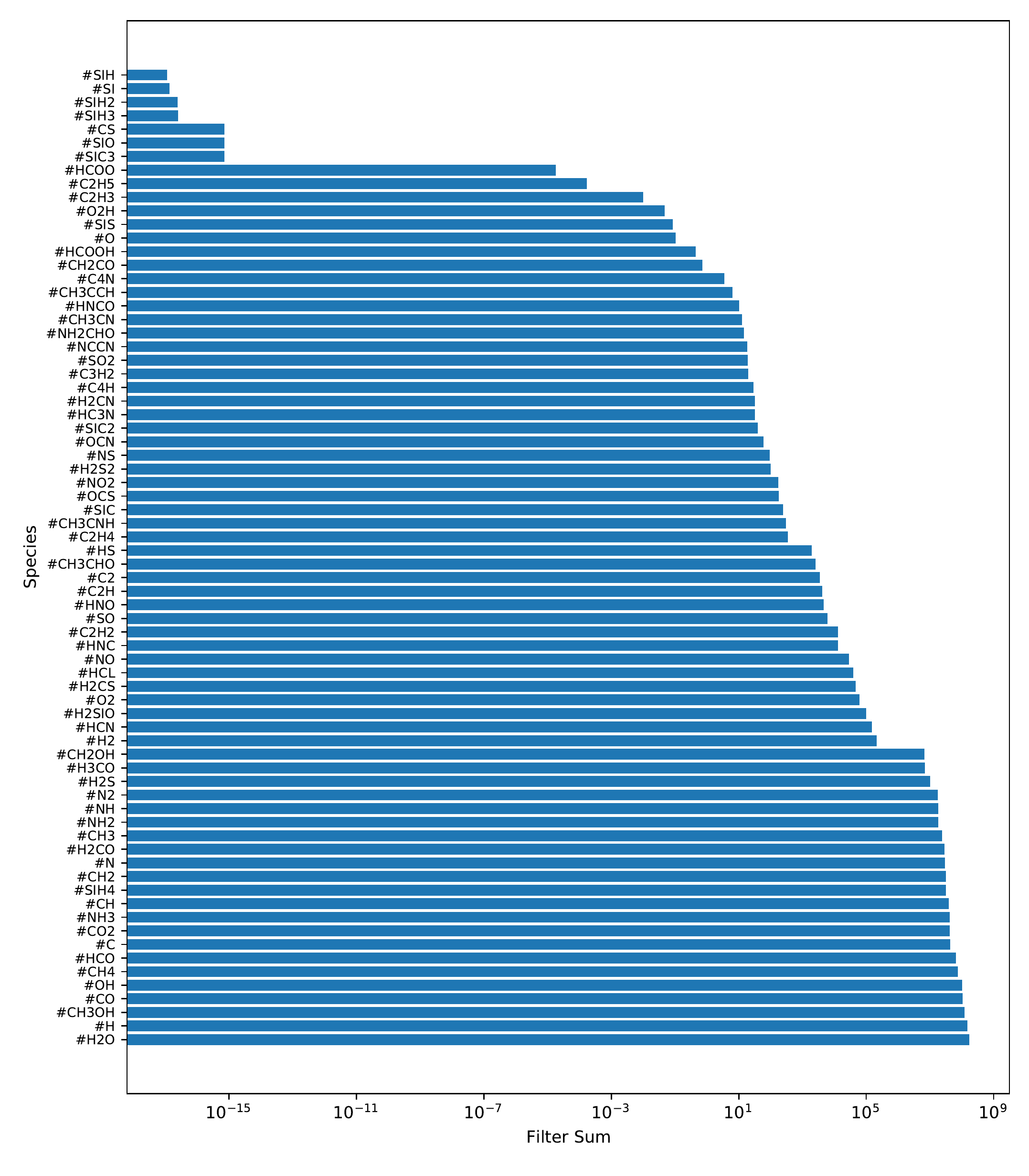}
    \caption{Bar chart showing the filter sums for each species in ascending order. Species with a larger filter sum should be prioritised for detection. Many of the species we observe are the intermediate species formed during the creation of the saturated species in Table \ref{abundance_table}. This indicates that understanding these intermediate products is essential to better constraining the binding energies of interest. We also note that many of the highest-ranked species have already been detected. This suggests that future observations should aim to improve the level of precision of these abundance measurements.}
    \label{sum_bar_chart}
\end{figure*}

\begin{figure*}
   \centering 
   \hspace*{-2cm}
    \includegraphics[width=2.5\columnwidth]{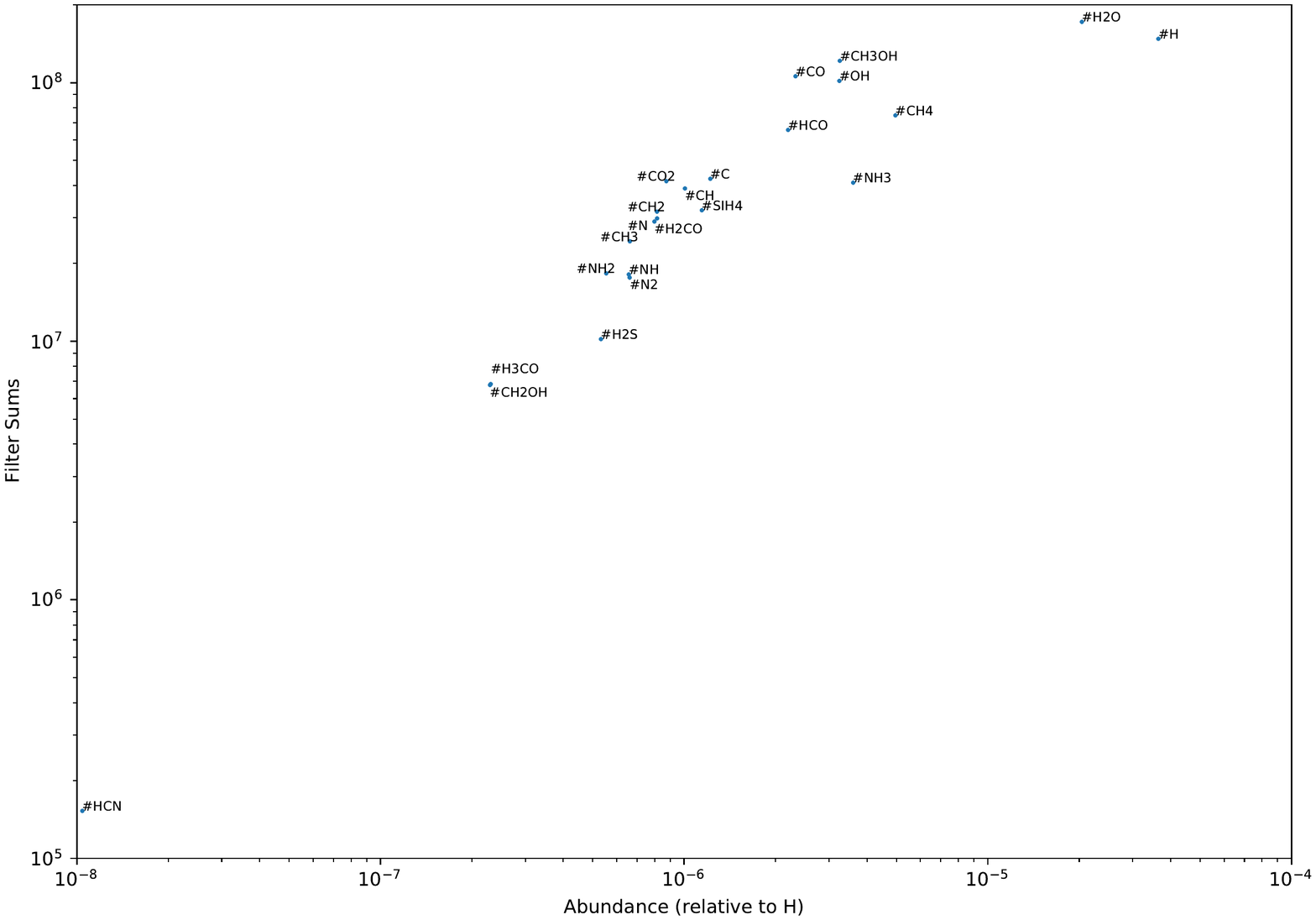}
    \caption{Scatter plot depicting filter sum against the predicted abundances when the MLE for binding energies are inserted into UCLCHEM. Given constraints on instrumental uncertainties, we should look to prioritise species that are not only important, as determined by their filter sums, but that can also be realistically detected. These include saturated species such as \#CH4, \#NH3, \#CO2 and \#H2O, but also their precursors.}
    \label{scatter_plot}
\end{figure*}


\begin{figure*}
    \includegraphics[width=2.3\columnwidth]{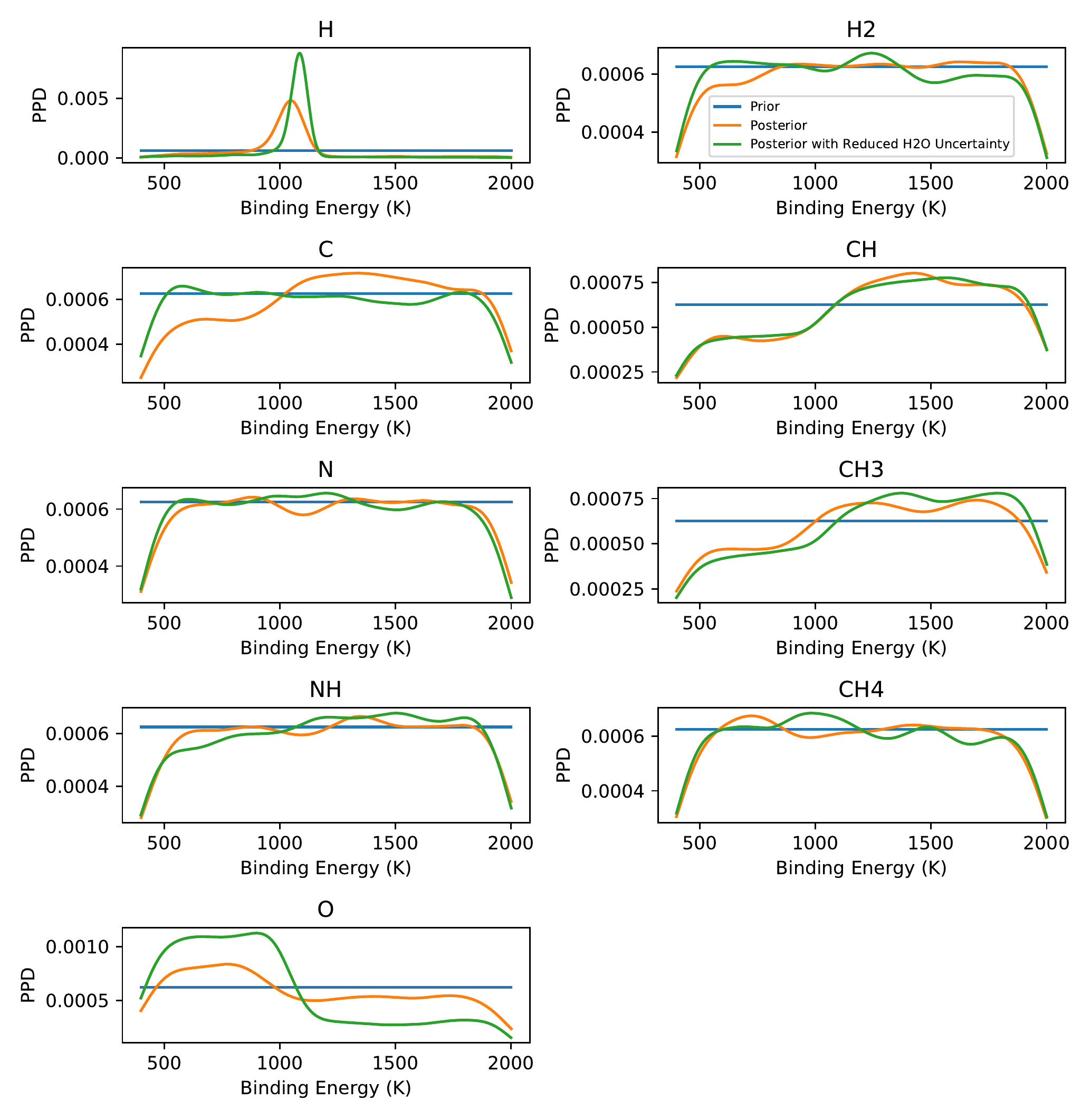}
    \caption{Marginalised posterior distributions of the binding energies of the diffusive species of interest. We also plot the prior distribution and the posterior distributions when when the uncertainty on water's abundance is reduced to $10^{-6}$. We observe that this has a significant effect on the marginalised posterior distributions of H and O, indicating that there is promise in improving the abundance measurements for species that have already been detected. }
    \label{binding_energy_plot_improved}
\end{figure*}

\subsection{Using MOPED}
We now look to use the MOPED algorithm to allow us to make predictions about which grain-surface species need to be detected in order to better constrain the posterior distribution. The maximum-likelihood estimate (MLE) from the inference was taken and partial derivatives taken around this point. It was found that near the MLE the partial derivatives of $\textbf{Y}$ with respect to the binding energies of C, NH, CH$_{4}$ and O were equal to the zero vector. This implies that for binding energies near the MLE, the reaction rates of the network are not sensitive to changes in the binding energies of these species. As such, these parameters were not included when calculating the filter values in the MOPED algorithm. 

Figure \ref{sum_bar_chart} shows the sum of the filters for all grain-surface species. The greater the filter sum, the more important it is to detect that molecule. Additionally, one must also consider the likely abundance of each species, as the species will only be observable in the ices if its abundance is above some minimum threshold. We therefore believe that future ice observations should prioritise species that have a high filter sum as well as a high abundance. In order to provide estimates of the abundances, we inserted the maximum-likelihood estimator values for the binding energy, $\boldsymbol{E_{ML}}$, into UCLCHEM and obtained the fitted abundances for all the species. Figure \ref{scatter_plot} is a scatter plot of the filter sum values against the abundances for each species. From this plot, we are able to identify high-importance species that are also likely to be detectable in the ices. However, one needs to also account for which species are realistic targets from a chemical point of view. This is discussed in the next subsection.

\subsection{Observational Implications}
The MOPED analysis has resulted in a clear ranking of which species should be targeted in future ice observations. This ranking is shown in Figure \ref{sum_bar_chart}. Of course we note that many of these species have very low abundances and others are difficult to detect in absorption. Diatomic molecules, atomic species and all radicals except CO will be neglected in our considerations of which species to consider. 

We briefly return to the issue of the network's reliability which was first discussed in Section~\ref{sec:network}. Whilst one can be confident in the abundances of CH$_4$, H$_{2}$CO, CH$_3$OH and H$_2$O as their networks are experimentally derived \citep{CH3OH, CO2, Chuang, CH4}, other species should be viewed more skeptically. This is particularly the case for sulphur. Many works indicate sulfur may primarily be locked in other forms \citep{Vidal,sulphur_woods}. It may be that the sulphur reaction network is incomplete. Most concerning is H$_2$S which the model suggests is the primary sulphur reservoir on the grains. Observations of ices have never detected H$_2$S but have instead provided upper limits of $\sim$10$^{-6}$\citep{Boogert}. The most likely value of the H$_2$S abundance derived here is lower than this limit and so it may be correct. However, there are other species in the network such as CS whose surface chemistry is not well-understood \citep{sulphur_woods}.  Taking this into consideration, it could be argued that observers should instead target species such as H$_{2}$CO or HCN which have similar filter sums and more reliable networks despite their lower predicted abundances.


There is much to be gained from obtaining more precise measurements for the abundances of species listed in Table \ref{abundance_table}. All of these species except for HCOOH and NH$_{4}^{+}$ have high filter sums and high abundances in the fitted model. However, the uncertainties on the measured abundances are often 50\% of the measured value. Our MOPED analysis shows that it would actually be much more valuable to determine these abundances to a smaller degree of uncertainty than it would be to measure the abundance of new species. To demonstrate, the effect of reducing the uncertainties on the abundances, we redid the Bayesian analysis, but reduced the uncertainty on water's abundance to $10^{-6}$. Figure \ref{binding_energy_plot_improved} shows the resulting binding energy posteriors. We observe significant changes in the posterior distributions for H and O. This suggests that there is much promise in improving the measured ice abundances for those molecules. Many of the absorption band profiles for these species are in the wavelength range of JWST, but especially in the 5-8 $\mu$m range that will have higher resolution compared to Spitzer \citep{Boogert}. This is promising as it is certain that H$_2$O and the other abundant species can be observed and telescope time simply needs to be dedicated to further constraining their abundances.

The infrared absorption profile of HCN has been studied recently in a laboratory setting \citep{HCN_paper}. Values for selected IR absorptions of amorphous HCN at 10 K were given including the C-H stretch (3.19 $\mu$m), the C$\equiv$N stretch (4.75 $\mu$m) and the HCN bend (12.12 $\mu$m). These as well as the combination and overtone features are well within the range of wavelengths that JWST will consider. As such, this would be a viable target molecule. 

While there might be some uncertainties relating to the sulphur network, H$_2$S has indeed a high fitted abundance as well as a high filter sum, hence it could potentially remain a target. There currently only exists an upper limit for the abundance of H$_2$S which was noted in \cite{H2S_first_identification}. This work identified a an S-H stretch mode at 3.925 $\mu$m, with \citet{H2S_infrared_spectroscopy} identifying an S-H stretching overtone mode at 1.982 $\mu$m.

SiH$_4$ is known to have several modes in the range 2.21 - 11.32 $\mu$m range \citep{SiH4, SiH4_other_paper}. These are all within the range that will be considered by JWST. 

H$_2$CO has its C$=$O stretching mode at around 5.8 $\mu$m, but this region is also host to other species with a C$=$O bond such as acetaldehyde, formic acid and formamide \citep{H2CO_paper, acetaldehyde}. It is thought to have another feature at 3.46 $\mu$m, which is however considerably weaker \citep{H2CO_paper}. It is for this reason that JWST's increased resolution in the 5-8 $\mu$m region would prove useful in separating out the various components.




\section{Conclusion}
In this work, we have utilised the MOPED algorithm to identify the species that would best constrain binding energies. Bayesian inference was found to result in poorly-constrained marginalised posterior distributions for the binding energies. This was due to the lack of enough sufficiently constraining data. The MOPED algorithm allowed us to determine which ice species should be prioritised for future ice observations in such a way that they would further constrain the posteriors. By then considering which species in the fitted model have the highest filter sums as well as the largest abundances, we come up with a list of species that should be targeted. These species are H$_2$O, CO$_2$, NH$_3$, CH$_4$, CO, CH$_3$OH, H$_2$CO, HCN, H$_2$S. While some of these species have not been detected, some of them have, which suggests that more precise measurements of these species is necessary. We also comment on which features of each species are likely to appear in the wavelength range considered by JWST. 

There are some limitations to this work. While our chemical network is for the most part reliable and reflects the current  understanding in the literature, there are still some uncertainties relating to particular species, such as sulphur. As such, if detecting sulphur species were a priority for future observations, then more work would need to be done to be completely confident of the sulphur network. 

Finally, one assumption that is made is that any species that will be detected will have the same level of uncertainty. This might not necessarily be true. The MOPED algorithm will favour species that have a strong dependence on the parameters, but also those which are low in variance. We have made use of the former, but not the latter in this work. For now, the results of this work are a proof-of-concept of the utility of the MOPED algorithm for this task.

\section*{Acknowledgements}
The authors thank the referee for their constructive
comments that greatly improved this work.J. Heyl is funded by an STFC studentship in Data-Intensive Science (grant number ST/P006736/1). This work was also supported by European Research Council (ERC) Advanced Grant MOPPEX 833460. S. Viti acknowledges support from the European Union’s Horizon 2020 research and innovation programme under the Marie Skłodowska-Curie grant agreement No 811312 for the project ``Astro-Chemical Origins” (ACO).

\section*{Data Availability}

The data underlying this article are available in the article and in its online supplementary material.



\bibliographystyle{mnras}
\bibliography{references} 

\begin{thebibliography}{}
\makeatletter
\relax
\def\mn@urlcharsother{\let\do\@makeother \do\$\do\&\do\#\do\^\do\_\do\%\do\~}
\def\mn@doi{\begingroup\mn@urlcharsother \@ifnextchar [ {\mn@doi@}
  {\mn@doi@[]}}
\def\mn@doi@[#1]#2{\def\@tempa{#1}\ifx\@tempa\@empty \href
  {http://dx.doi.org/#2} {doi:#2}\else \href {http://dx.doi.org/#2} {#1}\fi
  \endgroup}
\def\mn@eprint#1#2{\mn@eprint@#1:#2::\@nil}
\def\mn@eprint@arXiv#1{\href {http://arxiv.org/abs/#1} {{\tt arXiv:#1}}}
\def\mn@eprint@dblp#1{\href {http://dblp.uni-trier.de/rec/bibtex/#1.xml}
  {dblp:#1}}
\def\mn@eprint@#1:#2:#3:#4\@nil{\def\@tempa {#1}\def\@tempb {#2}\def\@tempc
  {#3}\ifx \@tempc \@empty \let \@tempc \@tempb \let \@tempb \@tempa \fi \ifx
  \@tempb \@empty \def\@tempb {arXiv}\fi \@ifundefined
  {mn@eprint@\@tempb}{\@tempb:\@tempc}{\expandafter \expandafter \csname
  mn@eprint@\@tempb\endcsname \expandafter{\@tempc}}}

\bibitem[\protect\citeauthoryear{{Belloche} et~al.,}{{Belloche}
  et~al.}{2017}]{Belloche_2017}
{Belloche} A.,  et~al., 2017, \mn@doi [\aap] {10.1051/0004-6361/201629724},
  \href {https://ui.adsabs.harvard.edu/abs/2017A&A...601A..49B} {601, A49}

\bibitem[\protect\citeauthoryear{{Boogert}}{{Boogert}}{2016}]{Boogert_conference}
{Boogert} A.~C.~A.,  2016, \mn@doi [IAU Focus Meeting]
  {10.1017/S174392131600315X}, \href
  {https://ui.adsabs.harvard.edu/abs/2016IAUFM..29A.317B} {29A, 317}

\bibitem[\protect\citeauthoryear{Boogert, Gerakines  \& Whittet}{Boogert
  et~al.}{2015}]{Boogert}
Boogert A.~A.,  Gerakines P.~A.,   Whittet D.~C.,  2015, \mn@doi [Annual Review
  of Astronomy and Astrophysics] {10.1146/annurev-astro-082214-122348}, 53, 541

\bibitem[\protect\citeauthoryear{{Buchner}}{{Buchner}}{2016}]{Buchner1}
{Buchner} J.,  2016, \mn@doi [Statistics and Computing]
  {10.1007/s11222-014-9512-y}, \href
  {https://ui.adsabs.harvard.edu/abs/2016S&C....26..383B} {26, 383}

\bibitem[\protect\citeauthoryear{{Buchner}}{{Buchner}}{2019}]{Buchner2}
{Buchner} J.,  2019, \mn@doi [\pasp] {10.1088/1538-3873/aae7fc}, \href
  {https://ui.adsabs.harvard.edu/abs/2019PASP..131j8005B} {131, 108005}

\bibitem[\protect\citeauthoryear{{Buchner}}{{Buchner}}{2021}]{UltraNest}
{Buchner} J.,  2021, \mn@doi [The Journal of Open Source Software]
  {10.21105/joss.03001}, \href
  {https://ui.adsabs.harvard.edu/abs/2021JOSS....6.3001B} {6, 3001}

\bibitem[\protect\citeauthoryear{{Caselli} \& {Ceccarelli}}{{Caselli} \&
  {Ceccarelli}}{2012}]{Caselli}
{Caselli} P.,  {Ceccarelli} C.,  2012, \mn@doi [\aapr]
  {10.1007/s00159-012-0056-x}, \href
  {https://ui.adsabs.harvard.edu/abs/2012A&ARv..20...56C} {20, 56}

\bibitem[\protect\citeauthoryear{{Chang}, {Cuppen}  \& {Herbst}}{{Chang}
  et~al.}{2007}]{Chang}
{Chang} Q.,  {Cuppen} H.~M.,   {Herbst} E.,  2007, \mn@doi [\aap]
  {10.1051/0004-6361:20077423}, \href
  {https://ui.adsabs.harvard.edu/abs/2007A&A...469..973C} {469, 973}

\bibitem[\protect\citeauthoryear{{Chuang}, {Fedoseev}, {Ioppolo}, {van
  Dishoeck}  \& {Linnartz}}{{Chuang} et~al.}{2016}]{Chuang}
{Chuang} K.~J.,  {Fedoseev} G.,  {Ioppolo} S.,  {van Dishoeck} E.~F.,
  {Linnartz} H.,  2016, \mn@doi [\mnras] {10.1093/mnras/stv2288}, \href
  {https://ui.adsabs.harvard.edu/abs/2016MNRAS.455.1702C} {455, 1702}

\bibitem[\protect\citeauthoryear{{Fathe}, {Holt}, {Oxley}  \&
  {Pursell}}{{Fathe} et~al.}{2006}]{H2S_infrared_spectroscopy}
{Fathe} K.,  {Holt} J.~S.,  {Oxley} S.~P.,   {Pursell} C.~J.,  2006, \mn@doi
  [Journal of Physical Chemistry A] {10.1021/jp0634104}, \href
  {https://ui.adsabs.harvard.edu/abs/2006JPCA..11010793F} {110, 10793}

\bibitem[\protect\citeauthoryear{{Fedoseev}, {Chuang}, {van Dishoeck},
  {Ioppolo}  \& {Linnartz}}{{Fedoseev} et~al.}{2016}]{Fedoseev_2016}
{Fedoseev} G.,  {Chuang} K.~J.,  {van Dishoeck} E.~F.,  {Ioppolo} S.,
  {Linnartz} H.,  2016, \mn@doi [\mnras] {10.1093/mnras/stw1028}, \href
  {https://ui.adsabs.harvard.edu/abs/2016MNRAS.460.4297F} {460, 4297}

\bibitem[\protect\citeauthoryear{{Ferrero}, {Zamirri}, {Ceccarelli}, {Witzel},
  {Rimola}  \& {Ugliengo}}{{Ferrero} et~al.}{2020}]{Ferrero}
{Ferrero} S.,  {Zamirri} L.,  {Ceccarelli} C.,  {Witzel} A.,  {Rimola} A.,
  {Ugliengo} P.,  2020, \mn@doi [\apj] {10.3847/1538-4357/abb953}, \href
  {https://ui.adsabs.harvard.edu/abs/2020ApJ...904...11F} {904, 11}

\bibitem[\protect\citeauthoryear{{Fuchs}, {Cuppen}, {Ioppolo}, {Romanzin},
  {Bisschop}, {Andersson}, {van Dishoeck}  \& {Linnartz}}{{Fuchs}
  et~al.}{2009}]{CH3OH}
{Fuchs} G.~W.,  {Cuppen} H.~M.,  {Ioppolo} S.,  {Romanzin} C.,  {Bisschop}
  S.~E.,  {Andersson} S.,  {van Dishoeck} E.~F.,   {Linnartz} H.,  2009,
  \mn@doi [\aap] {10.1051/0004-6361/200810784}, \href
  {https://ui.adsabs.harvard.edu/abs/2009A&A...505..629F} {505, 629}

\bibitem[\protect\citeauthoryear{{Garrod} \& {Herbst}}{{Garrod} \&
  {Herbst}}{2006}]{Garrod2006}
{Garrod} R.~T.,  {Herbst} E.,  2006, \mn@doi [\aap]
  {10.1051/0004-6361:20065560}, \href
  {https://ui.adsabs.harvard.edu/abs/2006A&A...457..927G} {457, 927}

\bibitem[\protect\citeauthoryear{{Garrod} \& {Pauly}}{{Garrod} \&
  {Pauly}}{2011}]{GarrodandPauly}
{Garrod} R.~T.,  {Pauly} T.,  2011, \mn@doi [\apj]
  {10.1088/0004-637X/735/1/15}, \href
  {https://ui.adsabs.harvard.edu/abs/2011ApJ...735...15G} {735, 15}

\bibitem[\protect\citeauthoryear{{Garrod}, {Widicus Weaver}  \&
  {Herbst}}{{Garrod} et~al.}{2008}]{Garrod_2008}
{Garrod} R.~T.,  {Widicus Weaver} S.~L.,   {Herbst} E.,  2008, \mn@doi [\apj]
  {10.1086/588035}, \href
  {https://ui.adsabs.harvard.edu/abs/2008ApJ...682..283G} {682, 283}

\bibitem[\protect\citeauthoryear{{Gerakines}, {Yarnall}  \&
  {Hudson}}{{Gerakines} et~al.}{2022}]{HCN_paper}
{Gerakines} P.~A.,  {Yarnall} Y.~Y.,   {Hudson} R.~L.,  2022, \mn@doi [\mnras]
  {10.1093/mnras/stab2992}, \href
  {https://ui.adsabs.harvard.edu/abs/2022MNRAS.509.3515G} {509, 3515}

\bibitem[\protect\citeauthoryear{{Hasegawa}, {Herbst}  \& {Leung}}{{Hasegawa}
  et~al.}{1992}]{Hasegawa}
{Hasegawa} T.~I.,  {Herbst} E.,   {Leung} C.~M.,  1992, \mn@doi [\apjs]
  {10.1086/191713}, \href
  {https://ui.adsabs.harvard.edu/abs/1992ApJS...82..167H} {82, 167}

\bibitem[\protect\citeauthoryear{{He}, {Acharyya}  \& {Vidali}}{{He}
  et~al.}{2016}]{experimental_approach}
{He} J.,  {Acharyya} K.,   {Vidali} G.,  2016, \mn@doi [\apj]
  {10.3847/0004-637X/825/2/89}, \href
  {https://ui.adsabs.harvard.edu/abs/2016ApJ...825...89H} {825, 89}

\bibitem[\protect\citeauthoryear{{Heavens}, {Jimenez}  \& {Lahav}}{{Heavens}
  et~al.}{2000}]{Heavens}
{Heavens} A.~F.,  {Jimenez} R.,   {Lahav} O.,  2000, \mn@doi [\mnras]
  {10.1046/j.1365-8711.2000.03692.x}, \href
  {https://ui.adsabs.harvard.edu/abs/2000MNRAS.317..965H} {317, 965}

\bibitem[\protect\citeauthoryear{{Heavens}, {Sellentin}, {de Mijolla}  \&
  {Vianello}}{{Heavens} et~al.}{2017}]{Heavens2017}
{Heavens} A.~F.,  {Sellentin} E.,  {de Mijolla} D.,   {Vianello} A.,  2017,
  \mn@doi [\mnras] {10.1093/mnras/stx2326}, \href
  {https://ui.adsabs.harvard.edu/abs/2017MNRAS.472.4244H} {472, 4244}

\bibitem[\protect\citeauthoryear{Heavens, Sellentin  \& Jaffe}{Heavens
  et~al.}{2020}]{Heavens2020ExtremePhysics}
Heavens A.~F.,  Sellentin E.,   Jaffe A.~H.,  2020, \mn@doi [Monthly Notices of
  the Royal Astronomical Society] {10.1093/mnras/staa2589}, 498, 3440

\bibitem[\protect\citeauthoryear{Herbst \& van Dishoeck}{Herbst \& van
  Dishoeck}{2009}]{Herbst}
Herbst E.,  van Dishoeck E.~F.,  2009, \mn@doi [Annual Review of Astronomy and
  Astrophysics] {10.1146/annurev-astro-082708-101654}, 47, 427

\bibitem[\protect\citeauthoryear{{Heyl}, {Viti}, {Holdship}  \&
  {Feeney}}{{Heyl} et~al.}{2020}]{Heyl1}
{Heyl} J.,  {Viti} S.,  {Holdship} J.,   {Feeney} S.~M.,  2020, \mn@doi [\apj]
  {10.3847/1538-4357/abbeed}, \href
  {https://ui.adsabs.harvard.edu/abs/2020ApJ...904..197H} {904, 197}

\bibitem[\protect\citeauthoryear{{Heyl}, {Holdship}  \& {Viti}}{{Heyl}
  et~al.}{2022}]{Heyl2}
{Heyl} J.,  {Holdship} J.,   {Viti} S.,  2022, \mn@doi [\apj]
  {10.3847/1538-4357/ac6606}, \href
  {https://ui.adsabs.harvard.edu/abs/2022ApJ...931...26H} {931, 26}

\bibitem[\protect\citeauthoryear{{Holdship}, {Viti}, {Jim{\'e}nez-Serra},
  {Makrymallis}  \& {Priestley}}{{Holdship} et~al.}{2017}]{UCLCHEM}
{Holdship} J.,  {Viti} S.,  {Jim{\'e}nez-Serra} I.,  {Makrymallis} A.,
  {Priestley} F.,  2017, \mn@doi [\aj] {10.3847/1538-3881/aa773f}, \href
  {https://ui.adsabs.harvard.edu/abs/2017AJ....154...38H} {154, 38}

\bibitem[\protect\citeauthoryear{{Holdship}, {Jeffrey}, {Makrymallis}, {Viti}
  \& {Yates}}{{Holdship} et~al.}{2018}]{holdship}
{Holdship} J.,  {Jeffrey} N.,  {Makrymallis} A.,  {Viti} S.,   {Yates} J.,
  2018, \mn@doi [\apj] {10.3847/1538-4357/aae1fa}, \href
  {https://ui.adsabs.harvard.edu/abs/2018ApJ...866..116H} {866, 116}

\bibitem[\protect\citeauthoryear{{Ioppolo}, {van Boheemen}, {Cuppen}, {van
  Dishoeck}  \& {Linnartz}}{{Ioppolo} et~al.}{2011}]{CO2}
{Ioppolo} S.,  {van Boheemen} Y.,  {Cuppen} H.~M.,  {van Dishoeck} E.~F.,
  {Linnartz} H.,  2011, \mn@doi [\mnras] {10.1111/j.1365-2966.2011.18306.x},
  \href {https://ui.adsabs.harvard.edu/abs/2011MNRAS.413.2281I} {413, 2281}

\bibitem[\protect\citeauthoryear{{Kaiser} \& {Osamura}}{{Kaiser} \&
  {Osamura}}{2005a}]{SiH4}
{Kaiser} R.~I.,  {Osamura} Y.,  2005a, \mn@doi [\aap]
  {10.1051/0004-6361:20040305}, \href
  {https://ui.adsabs.harvard.edu/abs/2005A&A...432..559K} {432, 559}

\bibitem[\protect\citeauthoryear{{Kaiser} \& {Osamura}}{{Kaiser} \&
  {Osamura}}{2005b}]{SiH4_other_paper}
{Kaiser} R.~I.,  {Osamura} Y.,  2005b, \mn@doi [\apj] {10.1086/432156}, \href
  {https://ui.adsabs.harvard.edu/abs/2005ApJ...630.1217K} {630, 1217}

\bibitem[\protect\citeauthoryear{{Keane}, {Tielens}, {Boogert}, {Schutte}  \&
  {Whittet}}{{Keane} et~al.}{2001}]{H2CO_paper}
{Keane} J.~V.,  {Tielens} A.~G.~G.~M.,  {Boogert} A.~C.~A.,  {Schutte} W.~A.,
  {Whittet} D.~C.~B.,  2001, \mn@doi [\aap] {10.1051/0004-6361:20010936}, \href
  {https://ui.adsabs.harvard.edu/abs/2001A&A...376..254K} {376, 254}

\bibitem[\protect\citeauthoryear{{Makrymallis} \& {Viti}}{{Makrymallis} \&
  {Viti}}{2014}]{Antonios}
{Makrymallis} A.,  {Viti} S.,  2014, \mn@doi [\apj]
  {10.1088/0004-637X/794/1/45}, \href
  {https://ui.adsabs.harvard.edu/abs/2014ApJ...794...45M} {794, 45}

\bibitem[\protect\citeauthoryear{{McElroy}, {Walsh}, {Markwick}, {Cordiner},
  {Smith}  \& {Millar}}{{McElroy} et~al.}{2013}]{UMIST}
{McElroy} D.,  {Walsh} C.,  {Markwick} A.~J.,  {Cordiner} M.~A.,  {Smith} K.,
  {Millar} T.~J.,  2013, \mn@doi [\aap] {10.1051/0004-6361/201220465}, \href
  {https://ui.adsabs.harvard.edu/abs/2013A&A...550A..36M} {550, A36}

\bibitem[\protect\citeauthoryear{{Minissale}, {Dulieu}, {Cazaux}  \&
  {Hocuk}}{{Minissale} et~al.}{2016}]{Minissale_2016}
{Minissale} M.,  {Dulieu} F.,  {Cazaux} S.,   {Hocuk} S.,  2016, \mn@doi [\aap]
  {10.1051/0004-6361/201525981}, \href
  {https://ui.adsabs.harvard.edu/abs/2016A&A...585A..24M} {585, A24}

\bibitem[\protect\citeauthoryear{{Minissale} et~al.,}{{Minissale}
  et~al.}{2022}]{characteristic_vibration_frequency}
{Minissale} M.,  et~al., 2022, \mn@doi [ACS Earth and Space Chemistry]
  {10.1021/acsearthspacechem.1c00357}, \href
  {https://ui.adsabs.harvard.edu/abs/2022ESC.....6..597M} {6, 597}

\bibitem[\protect\citeauthoryear{{Penteado}, {Walsh}  \& {Cuppen}}{{Penteado}
  et~al.}{2017}]{Penteado}
{Penteado} E.~M.,  {Walsh} C.,   {Cuppen} H.~M.,  2017, \mn@doi [\apj]
  {10.3847/1538-4357/aa78f9}, \href
  {https://ui.adsabs.harvard.edu/abs/2017ApJ...844...71P} {844, 71}

\bibitem[\protect\citeauthoryear{{Qasim}, {Fedoseev}, {Chuang}, {He},
  {Ioppolo}, {van Dishoeck}  \& {Linnartz}}{{Qasim} et~al.}{2020}]{CH4}
{Qasim} D.,  {Fedoseev} G.,  {Chuang} K.~J.,  {He} J.,  {Ioppolo} S.,  {van
  Dishoeck} E.~F.,   {Linnartz} H.,  2020, \mn@doi [Nature Astronomy]
  {10.1038/s41550-020-1054-y}, \href
  {https://ui.adsabs.harvard.edu/abs/2020NatAs...4..781Q} {4, 781}

\bibitem[\protect\citeauthoryear{{Quan}, {Herbst}, {Osamura}  \&
  {Roueff}}{{Quan} et~al.}{2010}]{Quan_2010}
{Quan} D.,  {Herbst} E.,  {Osamura} Y.,   {Roueff} E.,  2010, \mn@doi [\apj]
  {10.1088/0004-637X/725/2/2101}, \href
  {https://ui.adsabs.harvard.edu/abs/2010ApJ...725.2101Q} {725, 2101}

\bibitem[\protect\citeauthoryear{{Qu{\'e}nard}, {Jim{\'e}nez-Serra}, {Viti},
  {Holdship}  \& {Coutens}}{{Qu{\'e}nard} et~al.}{2018}]{Quenard}
{Qu{\'e}nard} D.,  {Jim{\'e}nez-Serra} I.,  {Viti} S.,  {Holdship} J.,
  {Coutens} A.,  2018, \mn@doi [\mnras] {10.1093/mnras/stx2960}, \href
  {https://ui.adsabs.harvard.edu/abs/2018MNRAS.474.2796Q} {474, 2796}

\bibitem[\protect\citeauthoryear{{Ruaud}, {Loison}, {Hickson}, {Gratier},
  {Hersant}  \& {Wakelam}}{{Ruaud} et~al.}{2015}]{Ruaud}
{Ruaud} M.,  {Loison} J.~C.,  {Hickson} K.~M.,  {Gratier} P.,  {Hersant} F.,
  {Wakelam} V.,  2015, \mn@doi [\mnras] {10.1093/mnras/stu2709}, \href
  {https://ui.adsabs.harvard.edu/abs/2015MNRAS.447.4004R} {447, 4004}

\bibitem[\protect\citeauthoryear{{Smith}}{{Smith}}{1991}]{H2S_first_identification}
{Smith} R.~G.,  1991, \mn@doi [\mnras] {10.1093/mnras/249.1.172}, \href
  {https://ui.adsabs.harvard.edu/abs/1991MNRAS.249..172S} {249, 172}

\bibitem[\protect\citeauthoryear{{Song} \& {K{\"a}stner}}{{Song} \&
  {K{\"a}stner}}{2016}]{Song2016}
{Song} L.,  {K{\"a}stner} J.,  2016, \mn@doi [Physical Chemistry Chemical
  Physics (Incorporating Faraday Transactions)] {10.1039/C6CP05727F}, \href
  {https://ui.adsabs.harvard.edu/abs/2016PCCP...1829278S} {18, 29278}

\bibitem[\protect\citeauthoryear{{Tegmark}, {Taylor}  \& {Heavens}}{{Tegmark}
  et~al.}{1997}]{TTH}
{Tegmark} M.,  {Taylor} A.~N.,   {Heavens} A.~F.,  1997, \mn@doi [\apj]
  {10.1086/303939}, \href
  {https://ui.adsabs.harvard.edu/abs/1997ApJ...480...22T} {480, 22}

\bibitem[\protect\citeauthoryear{{Terwisscha van Scheltinga}, {Marcandalli},
  {McClure}, {Hogerheijde}  \& {Linnartz}}{{Terwisscha van Scheltinga}
  et~al.}{2021}]{acetaldehyde}
{Terwisscha van Scheltinga} J.,  {Marcandalli} G.,  {McClure} M.~K.,
  {Hogerheijde} M.~R.,   {Linnartz} H.,  2021, \mn@doi [\aap]
  {10.1051/0004-6361/202140723}, \href
  {https://ui.adsabs.harvard.edu/abs/2021A&A...651A..95T} {651, A95}

\bibitem[\protect\citeauthoryear{{Vasyunin}, {Caselli}, {Dulieu}  \&
  {Jim{\'e}nez-Serra}}{{Vasyunin} et~al.}{2017}]{Vasyunin}
{Vasyunin} A.~I.,  {Caselli} P.,  {Dulieu} F.,   {Jim{\'e}nez-Serra} I.,  2017,
  \mn@doi [\apj] {10.3847/1538-4357/aa72ec}, \href
  {https://ui.adsabs.harvard.edu/abs/2017ApJ...842...33V} {842, 33}

\bibitem[\protect\citeauthoryear{{Vidal}, {Loison}, {Jaziri}, {Ruaud},
  {Gratier}  \& {Wakelam}}{{Vidal} et~al.}{2017}]{Vidal}
{Vidal} T. H.~G.,  {Loison} J.-C.,  {Jaziri} A.~Y.,  {Ruaud} M.,  {Gratier} P.,
    {Wakelam} V.,  2017, \mn@doi [\mnras] {10.1093/mnras/stx828}, \href
  {https://ui.adsabs.harvard.edu/abs/2017MNRAS.469..435V} {469, 435}

\bibitem[\protect\citeauthoryear{{Villadsen}, {Ligterink}  \&
  {Andersen}}{{Villadsen} et~al.}{2022}]{Villadsen}
{Villadsen} T.,  {Ligterink} N. F.~W.,   {Andersen} M.,  2022, arXiv e-prints,
  \href {https://ui.adsabs.harvard.edu/abs/2022arXiv220703906V} {p.
  arXiv:2207.03906}

\bibitem[\protect\citeauthoryear{{Wakelam}, {Loison}, {Mereau}  \&
  {Ruaud}}{{Wakelam} et~al.}{2017}]{Wakelam}
{Wakelam} V.,  {Loison} J.~C.,  {Mereau} R.,   {Ruaud} M.,  2017, \mn@doi
  [Molecular Astrophysics] {10.1016/j.molap.2017.01.002}, \href
  {https://ui.adsabs.harvard.edu/abs/2017MolAs...6...22W} {6, 22}

\bibitem[\protect\citeauthoryear{{Woods}, {Occhiogrosso}, {Viti},
  {Ka{\v{n}}uchov{\'a}}, {Palumbo}  \& {Price}}{{Woods}
  et~al.}{2015}]{sulphur_woods}
{Woods} P.~M.,  {Occhiogrosso} A.,  {Viti} S.,  {Ka{\v{n}}uchov{\'a}} Z.,
  {Palumbo} M.~E.,   {Price} S.~D.,  2015, \mn@doi [\mnras]
  {10.1093/mnras/stv652}, \href
  {https://ui.adsabs.harvard.edu/abs/2015MNRAS.450.1256W} {450, 1256}

\makeatother
\end{thebibliography}








\bsp	
\label{lastpage}
\end{document}